# Advanced Analysis Methods in High Energy Physics*

Pushpalatha C. Bhat

*Fermi National Accelerator Laboratory, P.O .Box 500, Batavia, IL 60510*

**Abstract.** During the coming decade, high energy physics experiments at the Fermilab Tevatron and around the globe will use very sophisticated equipment to record unprecedented amounts of data in the hope of making major discoveries that may unravel some of Nature's deepest mysteries. The discovery of the Higgs boson and signals of new physics may be around the corner. The use of advanced analysis techniques will be crucial in achieving these goals. I will discuss some of the novel methods of analysis that could prove to be particularly valuable for finding evidence of any new physics, for improving precision measurements and for exploring parameter spaces of theoretical models.

> *"A reasonable man adapts himself to the world.*
> *An unreasonable man persists to adapt the world to himself.*
> *So, all progress depends on the unreasonable one."*
> -Bernard Shaw.

## INTRODUCTION

The CDF and DØ experiments are preparing for a new and possibly a decade-long run at the upgraded Fermilab Tevatron. A new generation of accelerators and detectors are on the horizon. In the coming decade, we hope to discover the Higgs boson and find evidence of physics beyond the Standard Model (SM) such as Supersymmetry or Technicolor, or something completely unexpected. In order to achieve the goals of the high energy physics (HEP) community, I believe it is crucial that advanced and optimal data analysis methods be used both on-line and off-line [1,2].

In our quest to understand the universe, we continually experiment, analyze observations, interpret results and update our knowledge. In high energy physics, there was a time when we exposed nuclear emulsion targets to particle beams or took bubble chamber photographs of interactions and "recorded" data from scans off-line. In the not-so-distant past, we could afford the luxury of writing data to storage media based on simple interaction criteria and organize, reduce and analyze data completely off-line. But, as our knowledge-base increased, and as we began to address more complex problems, looking for extremely rare processes at higher beam energies and higher luminosities, it became necessary to sift through large amounts of data on-line before selected data were written out. Each new generation of experiments is more demanding than the previous in terms of data handling; the rates of interactions and the number of detector channels to be read-out often grow by orders of magnitude. Finding the signals of new physics become a veritable case of "finding needles in a hay-stack". So new paradigms and new technologies need to be identified, developed and adopted.

## INTELLIGENT DETECTORS

Today, data analysis in HEP experiments starts when a high energy event occurs. The electronic data from the detectors need to be transformed into useful "physics" information in real-time. One can envision that the calorimeter, for instance, can have "intelligence" close to its electronic read-out so that the clustering and energy measurements are readily available. Such information from different sub-detectors can be used to extract event features, such as number of tracks, high transverse momentum ($p_T$) objects and object identities. The features are then used to make a decision about whether the event should be recorded. So we need to build intelligent detectors and smart triggers! Feature extraction, classification or particle identification can be accomplished using algorithms implemented either in specialized hardware (neural network chips, for example) or in conventional hardware such as Field Programmable Gate Arrays (FPGAs) or Digital Signal Processors (DSPs). The H1 experiment at HERA, for example, has used neural network hardware in its Level-2 trigger. This has been operated successfully since 1996 and has been crucial for the rich physics results from H1. The project has been discussed in detail in these proceedings by Chris Kiesling [2]. Innovative data management on-line, and the use of smart algorithms encoded in trigger hardware would be beneficial in meeting the demands of data handling and analysis on-line. Use of expert and fuzzy logic systems in controls and monitoring of detector electronics is an area that has not received much attention and needs to be explored.

# OPTIMAL ANALYSIS METHODS

My golden rule for an optimal analysis is this:

*"Keep it simple.
As simple as possible.
Not any simpler."*          - Einstein

Most data analysis tasks such as charged particle tracking, particle identification and signal/background discrimination, fitting, parameter estimation, functional approximation (deriving various correction and rate functions) and data exploration, normally involve several measured quantities or 'feature variables". To obtain the best possible results it is necessary to make maximal use of information in the data and hence employ optimal multivariate methods of analysis [1,3].

The power of multivariate methods in discrimination tasks can be illustrated by the following simple example. In Fig.1, I have shown distributions of two observables x1 and x2 arising from two bi-variate Gaussians. One sees considerable overlap of the two classes in the one-dimensional projections (Fig. 1(a,b). But if one examines the data in 2-dimensions, one sees that the two classes of events are separable (Fig.1(c)). A Fisher linear discriminant, an appropriate linear combination of x1 and x2 plotted in Fig. 1(d), can provide a clear separation of the two classes.

In real life examples, decision boundaries between classes are more complicated and require the use of more sophisticated, more flexible, non-linear methods to calculate them. But, I want to emphasize that, as is true of all methods, it is important to make an appropriate choice of model complexity. A highly flexible model with lots of parameters will over-fit the data. This is illustrated by an example of polynomial fitting shown in Fig.2. The smooth curve is the parent function $f(x) = 0.5 + 0.4\sin(2\pi x)$. The data points are generated by adding random noise. Having either too few or too many parameters to fit the data yields a model that provides a poor representation of the underlying parent function.

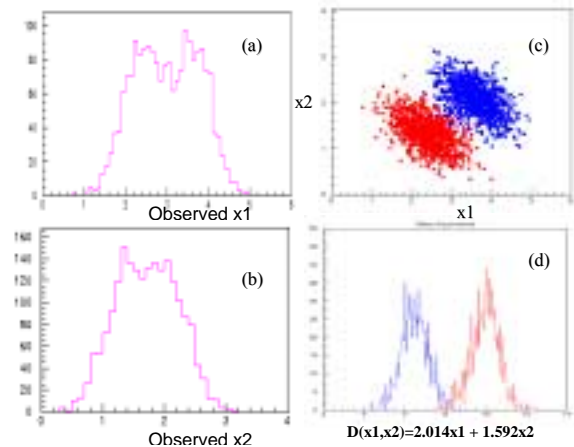

**FIGURE 1.** (a,b) Distributions of two hypothetical observables x1 and x2 arising from a mixture of two classes of events. (c) Original 2D distribution for the two classes of events and (d) Fisher linear discriminant that provides a mapping to 1-dimension in a way that cleanly separates the two classes.

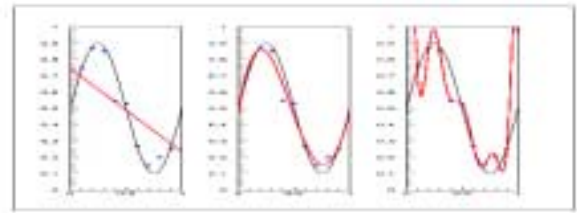

**FIGURE 2.** Results of fitting a data set (shown by points) with a $1^{st}$, $3^{rd}$, and $10^{th}$ order polynomial (plots from left to right). The generator function $f(x) = 0.5 + 0.4\sin(2\pi x)$ is superposed in each case.

Since event classification (or discrimination) is one of the most common tasks we deal with in high energy physics, I will concentrate on that topic in the rest of this section.

*Optimal discrimination* minimizes the probability of mis-classification. The traditional procedure of choosing and applying cuts on one event variable at a time is rarely optimal in that sense. However, given a set of event variables (denoted by a vector **x**), if correlations exist between them, optimal separation can *always* be achieved if one treats the variables in a fully multivariate manner. The optimal way to partition a multidimensional space populated by two classes of events 's' and 'b', for example, is to apply a cut on the ratio of the probabilities,

$$r(\mathbf{x}) = \frac{p(s|\mathbf{x})}{p(b|\mathbf{x})} = \frac{p(\mathbf{x}|s)p(s)}{p(\mathbf{x}|b)p(b)}, \qquad (1)$$

where $p(\mathbf{x}|s)$ and $p(\mathbf{x}|b)$ are the class conditional probabilities, that is, probability density functions for signal and background, respectively; $p(s)$ and $p(b)$ are the prior probabilities.

The posterior probability for the desired class 's' then becomes,

$$p(s|\mathbf{x}) = \frac{r}{1+r} = \frac{p(\mathbf{x}|s)p(s)}{p(\mathbf{x}|s)p(s) + p(\mathbf{x}|b)p(b)}. \qquad (2)$$

The discriminant '*r*' is called the Bayes discriminant. The problem of discrimination, then, mathematically reduces to that of calculating the Bayes discriminant $r(\mathbf{x})$ or the class conditional probabilities. I should note here that algorithms such as neural networks, interestingly, can directly yield the posterior probability $p(s/\mathbf{x})$.

In general, when many classes $C_k$ ($k=1,\ldots,N$) are present, the Bayes posterior probability is written as,

$$p(C_k|\mathbf{x}) = \frac{p(\mathbf{x}|C_k)p(C_k)}{\sum p(\mathbf{x}|C_k)p(C_k)}. \qquad (3)$$

The Bayes rule for classification is to assign the object to the class with highest posterior probability.

## PROBABILITY DENSITY ESTIMATION

I will briefly describe a few popular multivariate methods, most of which are probability density estimators.

## Histogramming

The problem of probability density estimation in principle can be solved quite simply! One would merely histogram the multivariate data **x** in **M** bins in each of the $d$ feature variables. The fraction of events (points) that fall within each bin yields a direct estimate of the density at the value of the feature vector **x**, say at the center of the bin. The bin width (and therefore the number of bins M) has to be chosen such that the structure in the density is not washed out (due to too few bins) and the density estimation is not too spiky (due to too many bins). The serious disadvantage of the histogramming method is that the total number of bins required grows like $\mathbf{M}^d$ (referred to as Bellman's curse of dimensionality). We would require a huge number of data points or else most of the bins would be empty leading to an estimated density of zero for those bins. The other issue is that the variables are generally correlated, and tend to be restricted to a sub-space of lower dimensionality, referred to as *intrinsic dimensionality*. Clearly, this method is inadequate for high dimensional data. There are better and more efficient methods for density estimation.

## Kernel-based Methods

These methods sample neighborhoods of data points to provide probability densities. Let us take the simple example of a hypercube of side $h$ as the kernel function in a $d$-dimensional space. The method consists of placing such a hypercube at each data point $\mathbf{x}_n$, counting the number of data points that fall within the hypercube and dividing that by the volume of the hypercube and the total number of data points, i.e.,

$$\tilde{p}(\mathbf{x}) = \frac{1}{N}\sum_{n=1}^{N}\frac{1}{h^d}H\left(\frac{\mathbf{x}-\mathbf{x}_n}{h}\right), \qquad (4)$$

where N is the total number of data points, and

$H(u)=1$ if **x** is in the hypercube, 0 otherwise.

The method is akin to histogramming, but with overlapping bins (hypercubes) this time placed around each data point. Smoother and more robust density estimates can be obtained by using smooth functional forms for the kernel function. A common choice is a multivariate Gaussian,

$$H(u) = \frac{1}{(2\pi)^{d/2}} \exp\left(-\frac{\|\mathbf{x}-\mathbf{x}^n\|^2}{2h^2}\right), \quad (5)$$

where the width of the Gaussian $h$ acts as a smoothing parameter to be chosen appropriately for the problem.

If the kernel functions satisfy,

$$H(u) \geq 0; \quad \int H(u)du = 1 \quad (6)$$

then, the estimator satisfies

$\widetilde{p}(\mathbf{x}) \geq 0$ and $\int \widetilde{p}(\mathbf{x})d\mathbf{x} = 1$, as required.

The PDE method[4], used at DØ in the measurement of the top quark mass using dilepton events [5], is an example of such a kernel-based method.

## K-Nearest Neighbor Method

In the kernel-based approach, the parameter $h$ is a constant and consequently the density estimation can be over-smoothed in some regions and spiky in some others. This problem is addressed in the K-nearest-neighbor approach. In this case, we place a kernel, say a hypersphere, at each data point $\mathbf{x}$ and instead of fixing its volume V and counting the number of data points that fall within it, we vary the volume (i.e., the radius of the hypersphere) until a fixed number of data points are within the volume. Then, the density is calculated as,

$$\widetilde{p}(\mathbf{x}) = \frac{K}{NV}. \quad (7)$$

A classification criterion can be directly obtained in the K-nearest-neighbor approach as follows: if there are $N_k$ points belonging to class $C_k$ and $N$ points in total, so that $\sum_k N_k = N$, then the class conditional probabilities can be written as

$$p(\mathbf{x}|C_k) = \frac{K_k}{N_k V}, \quad (8)$$

where $K_K$ is the number of points in volume V for class $C_K$.

The prior probability,

$$p(C_k) = \frac{N_K}{N}. \quad (9)$$

The Bayes posterior probability is;

$$p(C_k|\mathbf{x}) = \frac{p(\mathbf{x}|C_k)P(C_k)}{p(\mathbf{x})} = \frac{K_k}{K}. \quad (10)$$

This yields the following algorithm: a new feature vector $\mathbf{x}$ should be assigned to the class $C_k$ that has the most representatives in the volume of the hypersphere.

The contribution from Carli and Koblitz [6] at this workshop is an example of this method.

## Adaptive Mixtures

The method of adaptive mixtures (AM; also called the mixture model) is a variant of the Kernel-based approach where the density estimate is obtained by a linear combination of an adjustable number of basis functions or component densities $p(\mathbf{x}|j)$,

$$\widetilde{p}(\mathbf{x}) = \sum_{j=1}^{M} p(\mathbf{x}|j)p(j), \quad (11)$$

where M is typically far less than the number of points N, and the coefficients $p(j)$ are the mixing parameters. The most common functional form assumed for the component densities is a multivariate Gaussian,

$$p(\mathbf{x}|j) = \frac{1}{(2\pi)^{d/2}|\Sigma_j|^{1/2}} \times$$
$$\exp\left[-\tfrac{1}{2}(\mathbf{x}-\boldsymbol{\mu}_j)^T \Sigma_j^{-1} (\mathbf{x}-\boldsymbol{\mu}_j)\right] \quad (12)$$

where $\boldsymbol{\mu}_j$ is the mean and $\Sigma_j$ is the covariance matrix. The adaptive mixtures algorithm would incorporate rules for adding or deleting components and for adjusting $\boldsymbol{\mu}_j$ and $\Sigma_j$.

The mixture models or the method of mixtures have been used quite extensively in the statistical community. These traditional applications assume that the data came from a mixture of a given number of components, where as in AM this assumption is not made.

## Neural Networks

Even though the concepts of neural networks were inspired from biology, the algorithms have deep statistical underpinnings. Neural network algorithms have emerged as powerful and flexible methods for a variety of multivariate data analysis applications. Feed-forward neural networks, also known as multilayered perceptrons, are the most popular and widely used. The output of a feed-forward neural network trained by minimizing, for example, mean square error function, directly approximates the Bayesian posterior probability $p(s|\mathbf{x})$ (Eq. 2) [7] without the need to estimate the class-conditional probabilities separately. A schematic of a feed-forward neural network (NN) is shown in Fig. 3. Such networks provide a general framework for estimating non-linear functional mappings between a set of input variables $\mathbf{x} (\equiv (x_1, x_2, x_3,... x_k))$ and an output variable $O(\mathbf{x})$ (or a set of output variables) without requiring a prior mathematical description of how the output formally depends on the inputs. The mapping involves transforming the input variables with an arbitrary number of adaptive non-linear functions. The output in the simple example shown in Fig. 3 can be written as,

$$O(\mathbf{x}) = g\left(\sum_i w_j h_j + \theta_i\right) \equiv p(s|\mathbf{x}), \quad (13)$$

with $h_j = g\left(\sum w_{jk} + \theta_j\right),$

and where $g$ is a non-linear "activation" function normally taken as a logistic sigmoid

$$g(a) = \frac{1}{1+e^{-a}}. \quad (14)$$

These "hidden" transformation functions $g$, or more precisely the weights $w_j$ and $w_{jk}$ and the thresholds (not shown in the figure) $\theta_i$ and $\theta_j$ adapt themselves to the data as part of the "training" process of the neural network. The number of such parameters need to grow only as the complexity of the problem grows. The parameters are determined by minimizing an error function, usually the mean square error between the actual output $O^p$ and the desired (target) output $t^p$,

$$E = \frac{1}{2N_p}\sum_{p=1}^{N}\left(O^p - t^p\right)^2, \quad (15)$$

with respect to the parameters. Here, $p$ denotes a feature vector or pattern. The stochastic optimization algorithms used in learning enable the model to be improved a little bit for each data point in the training sample. The Bayes discriminant in terms of the NN output is

$$r(\mathbf{x}) = \frac{O(\mathbf{x})}{1-O(\mathbf{x})}. \quad (16)$$

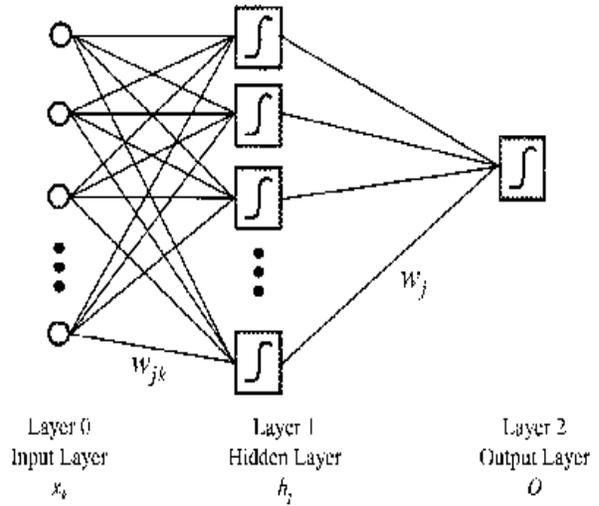

**FIGURE 3.** A schematic representation of a three layer feed-forward neural network.

Neural networks, apart from being universal approximators (i.e., they approximate probability densities or posterior probabilities to arbitrary accuracy), provide a very practical tool because of the relatively small computational times required in their training ( in a majority of applications in HEP). The fast convergence as well as the robustness in supervised learning of multilayer perceptrons are due to efficient and powerful algorithms developed in recent years.

Good generalization, that is good predictions for new inputs, is controlled by model complexity as we discussed in the example of polynomial curve fitting in the previous section. The traditional approaches used to control model complexity are *structure stabilization* (optimizing the size of the network) and *regularization.* In the former one starts with large networks and *prunes* connections or starts with small networks and adds units/neurons as necessary. In regularization, one penalizes complexity by adding a penalty term to the error function.

There are many new and sophisticated approaches to achieve good generalization. It is important to note here that the generalization error (g.e.) of an NN can be decomposed into the sum of the bias-squared ($b^2$) plus the variance ($\sigma^2$), i.e., the generalization error,
$$g.e. \equiv \sqrt{b^2 + \sigma^2}\ .$$

The goal is to minimize the g.e., that is, finding the best compromise between bias and variance. Ensembles of networks, such as committees or stacks, can be used to control bias and variance [8].

Bayesian learning of network parameters can in principle handle networks of arbitrarily high complexity without over-fitting. Bayesian networks also provide a rigorous way to assign errors to network predictions [8].

Aside from the MLP, there are other neural network types which are potentially useful in some applications in HEP. One of them is the self-organizing map (SOM). This is an unsupervised technique and appears to be an excellent tool for model-independent data exploration. It maps input space on to a low-dimensional (usually 2-D) regular grid that can be used to visualize and explore properties of the data. Given models for background processes, one could use it in a manner similar to the program "Sleuth" developed at DØ to search for new physics [9].

More detailed accounts and discussions of neural networks and other methods can be found in ref. [8]

## PHYSICS ANALYSIS EXAMPLES

I describe here a couple of example physics analyses to illustrate the power of multivariate methods.

### Top Quark Mass Measurement

The top quark mass measurement was one of the most important results from Run I of the Tevatron collider experiments. Since the DØ experiment did not have a silicon vertex detector (SVX) and used only soft muon tagging for b-jet identification, the b-tagging efficiency was only 20% in the lepton + ≥ 4-jets channel ($t\bar{t} \to Wb W\bar{b} \to l\nu bq\bar{q}\bar{b}$ process) compared to approximately 53% at CDF which had the ability to tag b-jets with its SVX.

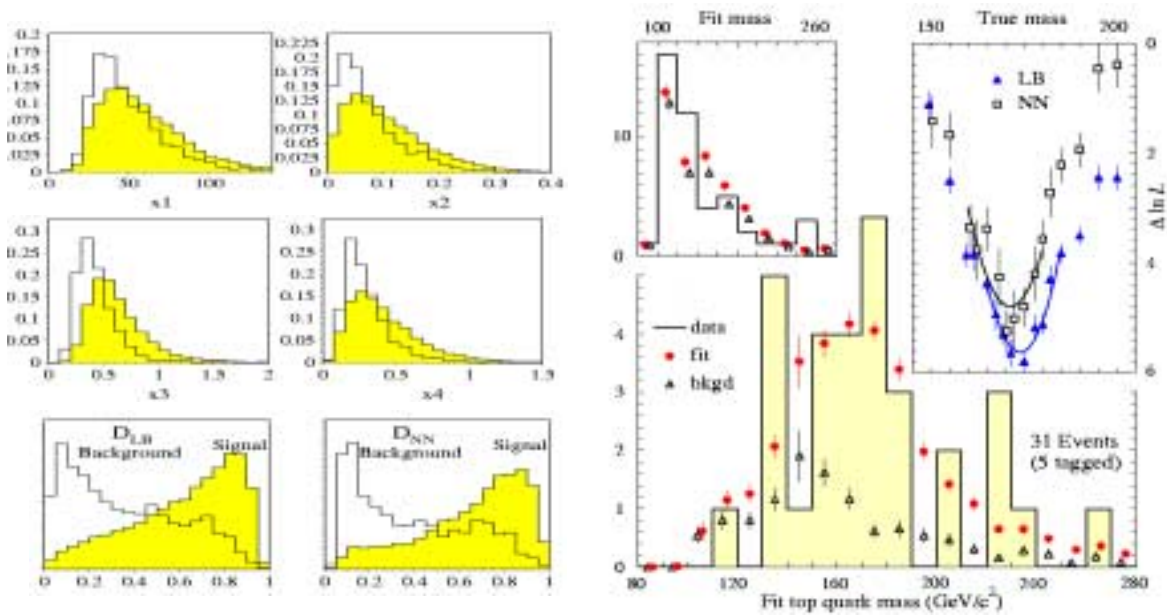

**FIGURE 4.** Distributions of discriminant variables $x_1, x_2, x_3, x_4$ (see [10] for definitions) and the final multivariate discriminant $D$ for signal (filled histograms) and background. All histograms are normalized to unity. (Right) The fitted mass distribution for events in the signal-rich sample. Left inset: The same for events in background–rich sample. Right inset: The relative log likelihood functions for the two methods. For details of the analysis, see ref. [11].

Nonetheless, DØ was able to measure the top quark mass with a precision approaching that of CDF, by using multivariate techniques for separating signal and background while minimizing the correlation of the selection with the top quark mass.

Two multivariate methods, (1) a modified log-likelihood technique (LB method) and (2) a feed forward neural network (NN method), were used to compute a signal probability $D \equiv p(top | \mathbf{x})$ for each event, given data $\mathbf{x}$. A likelihood fit, based on a Bayesian method [4], of the data to discrete sets of signal and background models in the $[p(top | \mathbf{x}), m_{fit}]$ plane was used to extract the top quark mass. ($m_{fit}$ is the fitted mass for each event from a kinematic fit to the $t\bar{t}$ hypothesis.) The distributions of variables and the results of the fits are shown in Fig. 4. Combining the results of the fits from the two methods, DØ measures $m_t = 173.3 \pm 5.6$ (stat) $\pm 5.5$ (syst) GeV/$c^2$ [11].

## Discovering the Higgs Boson

In the SM framework, a global fit to the electroweak precision data, including the directly measured top quark and W boson masses, yields a Higgs boson mass of $M_H = 107^{+67}_{-45}$ GeV/$c^2$ and a 95% C.L. upper limit of 225 GeV/$c^2$ [12]. In broad classes of supersymmetric (SUSY) theories, the mass $m_h$ of the lightest CP-even neutral Higgs boson $h$ is constrained to be less than 150 GeV/$c^2$ [13]. In the minimal supersymmetric SM (MSSM), $M_h < 130$ GeV/$c^2$ and there are tantalizing hints of a 115 GeV/$c^2$ Higgs boson from the recently completed LEP experiments [14]. These intriguing indications of a low-mass Higgs boson motivated studies of strategies that maximize the potential for its discovery at the upgraded Tevatron [15]. Our study of the Higgs discovery potential focused on a standard model Higgs boson in the mass range 90 GeV/$c^2$ < $M_H$ < 130 GeV/$c^2$ that would be produced via the processes,

$p\bar{p} \to WH \to l\nu bb, \ p\bar{p} \to ZH \to llbb, \nu\bar{\nu} b\bar{b}$.

The dominant backgrounds in these channels come from $Wb\bar{b}, WZ, t\bar{t}$ and single top processes. We have shown that a neural network analysis could yield a 5σ discovery for $100 \le M_H \le 130$ GeV/$c^2$ with only half the integrated luminosity needed for a conventional analysis. Fig. 5 shows the neural network distributions for signal Monte Carlo events with $M_H = 110$ GeV/$c^2$ compared with the specified backgrounds, for a set of seven input variables. (For details, see ref. 15). A plot of the required integrated luminosity for a 5σ observation is also shown in Fig. 5. For a 110 GeV/$c^2$ Higgs boson, if 10% systematic uncertainties are assumed, CDF and DØ would require about 13 fb$^{-1}$ for independent 5σ discovery. Our study shows that with 20 fb$^{-1}$, a 3-5σ observation of a neutral Higgs boson is possible at the Tevatron for masses with $M_H \le 130$ GeV/$c^2$.

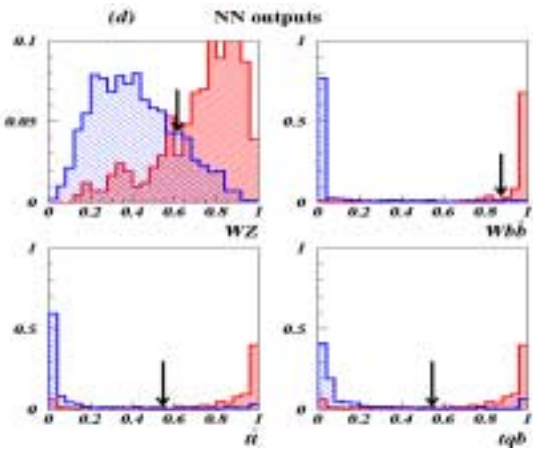 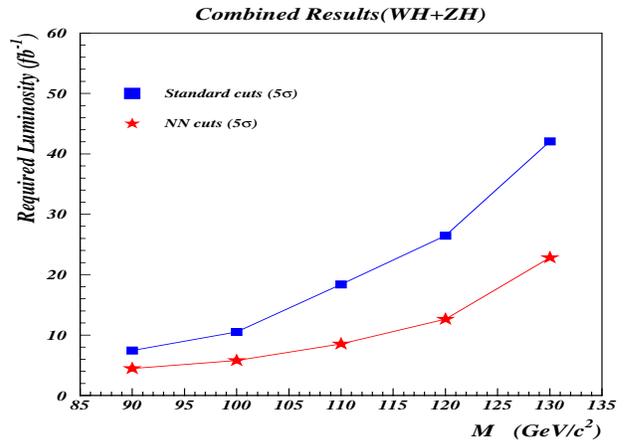

**FIGURE 5.** (Left) Neural network distributions for WH signal ($M_H = 110$ GeV/$c^2$; heavily shaded histograms) compared with backgrounds $Wb\bar{b}, WZ, t\bar{t}$ and single top. (Right) Comparison of the required integrated luminosities for a 5σ observation in the CDF and DØ experiments for NN and conventional cuts (WH and ZH channels combined).

# EXPLORING MODELS

Physicists are becoming increasingly convinced of the value of Bayesian reasoning as a powerful way of extracting information from data and of updating knowledge upon arrival of new data. The Bayesian approach provides a well-founded mathematical procedure to compute the conditional probability of a model (or a hypothesis) and therefore to do straight-forward and meaningful model comparisons. It also allows treatment of all uncertainties in a consistent manner. We have applied these ideas in two analyses (1) fitting binned data to one or more multi-source models [16] which was eventually used in the top quark mass measurement at DØ and (2) the extraction of the solar neutrino survival probability [17] as a function of neutrino energy, using data and solar neutrino model predictions. These practical applications illustrate the usefulness of Bayesian methods in data analysis.

The Bayesian approach provides a systematic way of extracting probabilistic information for each parameter of a model, say for example, a particular SUSY model, via marginalization over the remaining parameters. This probabilistic approach to model exploration could prove to be extremely fruitful. We are studying this approach in the search for supersymmetric Higgs boson predicted by the SO(10) model [18].

# CONCLUSIONS

The discovery of the Higgs boson and signals of new physics beyond the SM may be just around the corner in Run-II at the Fermilab Tevatron. Somewhat later, the experiments at the Large Hadron Collider at CERN will enable us to probe physics at the TeV scale. We[6]. are entering an exciting era with lots of optimism and hope. The physics pursuits are extremely challenging, even daunting!

Use of optimal analysis methods will have to become routine in order to achieve the high energy physics goals for the coming decade. These methods, particularly neural network techniques have already made an impact on discoveries and precision measurements and I believe that they will be the methods of choice for future analyses.

# ACKNOWLEDGMENTS

My research is supported in part by the U.S. Department of Energy under contract number DE-AC02-76CH03000.

momentum. $x_4 \equiv \Delta R_{jj}^{min} . E_T^{min} / (E_T^l + E_T)$ where $\Delta R_{jj}^{min}$ is the minimum $\Delta R$ (the distance between jets in $\eta - \varphi$ space) of the six pairs of four jets and $E_T^{min}$ is the smaller jet $E_T$ from the minimum $\Delta R$ pair.

___________________________